\documentclass[12pt,halfline,a4paper]{ouparticle}

\begin{document}

\title{Artificial Intelligence for Optimal Learning:\break A Comparative Approach towards AI-Enhanced Learning Environments}

\author{%
\name{Ananth Hariharan}
\address{University of Illinois Urbana-Champaign}
\email{ahari8@illinois.edu}}

\abstract{In the rapidly evolving educational landscape, the integration of technology has shifted from an enhancement to a cornerstone of educational strategy worldwide. This transition is propelled by advancements in digital technology, especially the emergence of artificial intelligence as a crucial tool in learning environments. This research project critically evaluates the impact of three distinct educational settings: traditional educational methods without technological integration, those enhanced by non-AI technology, and those utilising AI-driven technologies. This comparison aims to assess how each environment influences educational outcomes, engagement, pedagogical methods, and equity in access to learning resources, and how each contributes uniquely to the learning experience. The ultimate goal of this research is to synthesise the strengths of each model to create a more holistic educational approach. By integrating the personal interaction and tested pedagogical techniques of traditional classrooms, the enhanced accessibility and collaborative tools offered by non-AI technology, and the personalised, adaptive learning strategies enabled by AI-driven technologies, education systems can develop richer, more effective learning environments. This hybrid approach aims to leverage the best elements of each setting, thereby enhancing educational outcomes, engagement, and inclusiveness, while also addressing the distinct challenges and limitations inherent in each model. The intention is to create an educational framework deeply attentive to the diverse needs of students, ensuring equitable access to high-quality education for all.
}

\date{}


\maketitle

\section{Introduction}
\subsection{Background}
The role of technology in education has evolved significantly, transitioning from a supplementary tool to a fundamental component of educational strategies globally. This shift has been driven by rapid advancements in digital technologies, including the increasingly pivotal role of artificial intelligence (AI) in learning environments. These technologies are not merely adjuncts but are central to the structuring and delivery of educational content. By integrating sophisticated digital tools, educational settings are transformed, offering diverse learning experiences that cater to individual needs. This evolution is critical as it reflects a broader societal shift towards digital fluency and is integral in preparing students for a technology-driven world. The impact of these technologies is profound, affecting not only pedagogical methods but also accessibility, engagement, and educational outcomes across various types of learning environments.

This topic is significant due to the transformative potential of technology, especially AI, in education, which promises to redefine teaching and learning parameters. As digital tools advance, understanding their impact on educational outcomes is crucial, particularly in a world where educational disparities are stark. Moreover, the COVID-19 pandemic has accelerated the adoption of digital learning tools, giving time-tested insights into the role of technology. Finally, the emergence of publicly available access to large language models (LLMs) functioning as generative artificial intelligence (GenAI) tools makes this research timely and relevant for educational policy and practice.

\subsection{Research Scope}

The research encompasses several dimensions, each of which aims to target different parts of the educational landscape.

\begin{enumerate}
\item
Examining academic performance and learning outcomes across different educational setups.

\item
Exploring how technologies influence students' engagement levels.

\item
Investigating the adaptation of instructional strategies with technology integration.

\item
Assessing the digital divide and equitable access to technology-enhanced education.
\end{enumerate}

This research examines three pivotal models of education, reflecting the technological shifts across different generations – from Pre-Generation Y, through Generations Y and Z, to Post-Generation Z. In the Pre-Generation Y era, while technology was present in education, the limited use of personal devices characterised it as predominantly technology-free (Tapscott, 1998). The turn of the millennium marked a revolutionary shift as personal devices became integral to education, initially with laptops and tablets and expanding to include mobile phones, virtual reality, and augmented reality systems (Prensky, 2001; Oblinger \& Oblinger, 2005). This transformation signifies a major leap in the integration of technology within educational settings. In the post-pandemic landscape, advancements in artificial intelligence and the widespread adoption of remote learning have further revolutionised educational practices, offering unprecedented levels of technological integration (Williamson, 2021). Each stage of this educational evolution presents unique benefits and challenges. A balanced approach that integrates the best aspects of each era is crucial to ensuring that the educational system maintains quality, accessibility, and impact for all students (Selwyn, 2016).

This research endeavour will contribute to a nuanced understanding of how technology transforms educational environments. By comparing no-tech, non-AI tech, and AI-enhanced educational structures, this study will provide insights that can guide educators, policymakers, and technologists in making informed decisions about technology integration in education. The findings could also offer strategic directions for future technological deployments in educational settings, ensuring they are effective, equitable, and supportive of all learners.

\section{Theoretical Framework}

\subsection{Technology-Free Learning}

Traditional classrooms, defined by direct teacher-student interactions without the aid of digital tools, have long formed the cornerstone of educational systems worldwide. This model, which emphasises personal interaction and traditional teaching methods, is renowned for its ability to foster a learning environment that enhances social learning and strengthens student-teacher relationships (Laurillard, 2002). The emphasis on face-to-face communication and direct feedback supports an educational dynamic that is deeply relational and intuitively responsive, attributes that are often cited as foundational to effective educational practices (Alexander, 2004).

\subsubsection{Limitations in Personalisation and Engagement}

Despite their strengths, traditional classrooms often face limitations in providing personalised learning experiences that modern educational technologies can facilitate (Bates, 2015). In environments where instruction is delivered without digital tools, teachers may struggle to offer bespoke support to each student due to time constraints and the fixed pace of curriculum delivery. This one-size-fits-all approach can hinder the ability of some students to progress at an optimal rate, potentially exacerbating educational disparities (Means, 2010).

Moreover, the lack of digital tools can also impede the dynamic and interactive learning experiences that engage today’s learners. In the digital age, the engagement of students can be significantly enhanced by multimedia content and interactive learning platforms that cater to a range of learning styles and preferences. Traditional methods, while robust, often cannot replicate the engagement levels that technology-enhanced learning environments provide, particularly in disciplines that benefit from visual and interactive materials (Mayer, 2009).

Another critical aspect where traditional classrooms may fall short is in preparing students for a workforce increasingly dominated by digital skills. As digital literacy becomes as essential as traditional literacy, reliance solely on non-digital teaching methods might not suffice to equip students with the necessary skills to thrive in technologically advanced job markets (Voogt \& Roblin, 2012).

\subsubsection{Potential Benefits and Necessity for Balance}

Despite these challenges, traditional classrooms offer a tested and potentially less distracting environment for learning, which is free from the interruptions and technical issues that can accompany digital tools. This setting can foster a strong sense of community and interpersonal growth among students and teachers, attributes that are sometimes diluted in online or hybrid models (Selwyn, 2016).

Continuing from the advantages of traditional classrooms in fostering community and interpersonal growth, the classical education system offers a well-rounded approach that balances Grammar, Logic, and Rhetoric to cultivate a comprehensive learning experience. This system is uniquely positioned to blend these benefits with the structured progression of intellectual development, emphasising a sequential learning process that nurtures students’ cognitive and social abilities effectively.

In the classical education model, Grammar represents the foundational stage where students acquire essential knowledge through direct instruction and rote learning. This stage focuses on absorbing facts, figures, and the basics of subjects, providing a solid foundation for more complex thinking. In traditional settings, this method benefits from the personal interaction and immediate feedback that teachers provide, which can be crucial for young learners (Hirsch, 1987). Such environments are conducive to learning basic skills without the distractions of modern technology, allowing students to focus deeply on the content.

As students advance, they move into the Logic stage, where reasoning and critical thinking skills are developed. Here, the focus shifts from mere knowledge acquisition to understanding and applying logic to various subjects. Discussions and debates play a vital role in this phase, helping students learn to reason clearly and critically. The personal interactions that are prominent in traditional classrooms are irreplaceable in developing these skills, as they allow for dynamic, real-time exchanges of ideas, fostering a deeper understanding and the ability to argue persuasively (Alexander, 2004).

The final stage of the classical education model is Rhetoric, where students learn to express their thoughts in coherent, persuasive language and apply the knowledge and logic skills they have acquired. This stage emphasises writing and speech, with students often encouraged to create persuasive arguments and presentations. The traditional classroom offers a unique advantage in this stage, as it provides numerous opportunities for public speaking and real-time feedback from peers and instructors, which are crucial for refining students' rhetorical abilities (Bates, 2015).

\begin{figure}[htp]
    \centering
    \includegraphics[width=10cm]{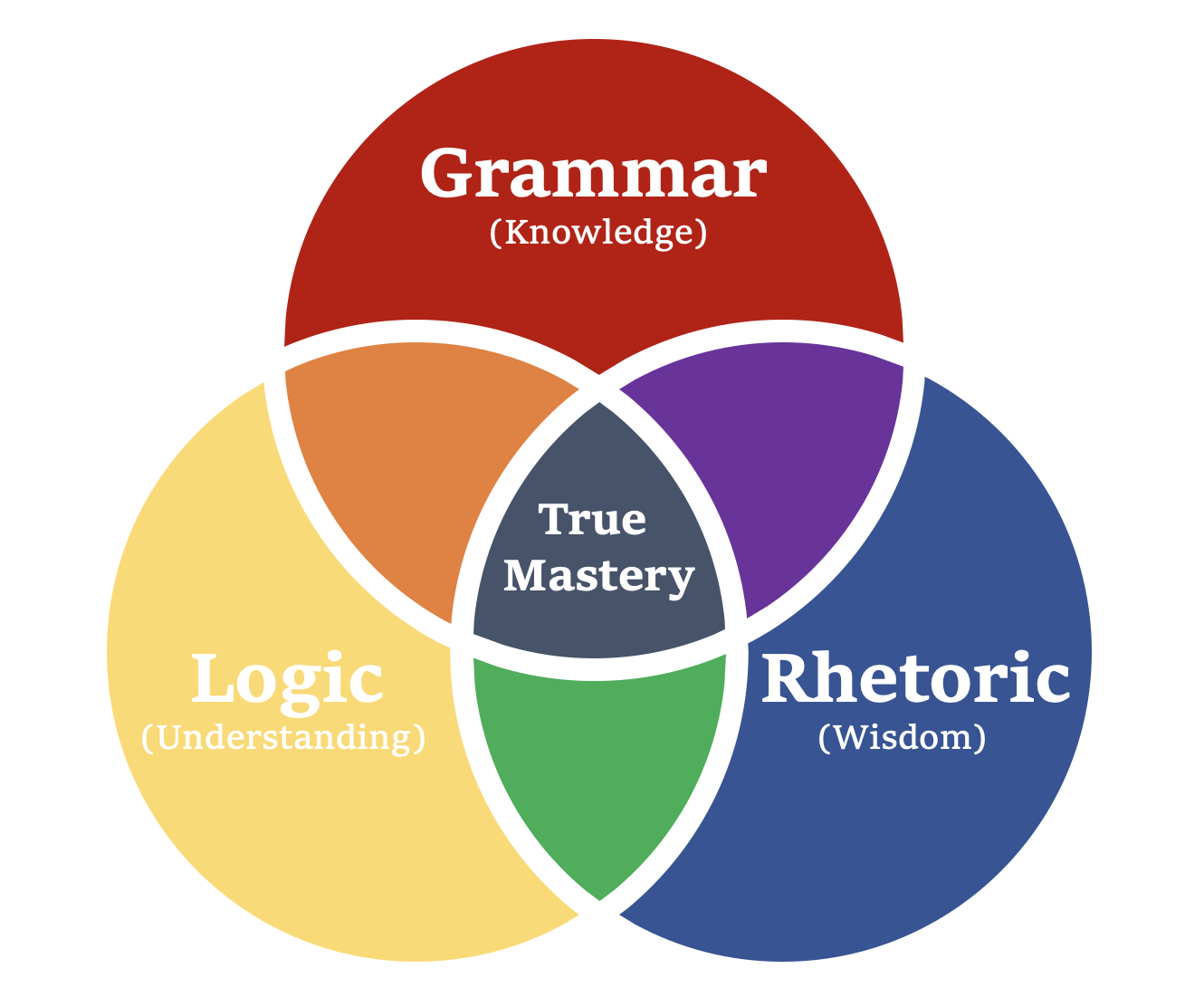}
    \caption{Classical Approach to Education}
    \label{fig:classic-venn}
\end{figure}

While the traditional classroom offers substantial benefits in teaching the trivium, integrating modern educational tools can enhance these foundational techniques. For instance, digital tools can provide access to a vast array of resources that enrich the Grammar stage, while AI-driven analytics can offer personalised insights into students’ understanding, aiding the Logic phase. Furthermore, technology like writing assistance tools and communication platforms can augment the Rhetoric stage by providing new forums and formats for expression.

Thus, while traditional classrooms play an indispensable role in classical education, there is potential for technology to support and enhance this model. By carefully integrating digital tools, educators can preserve the strengths of the classical system while embracing the benefits of technological advancements. This balanced approach can prepare students more comprehensively for the challenges of the modern world, ensuring they not only retain knowledge and think critically but also communicate effectively in an increasingly digital society.

\subsection{Conventional Technology in Education}

The integration of non-artificial intelligence (non-AI) technologies such as laptops, tablets, and mobile phones into educational settings has significantly transformed the landscape of teaching and learning. Unlike AI-driven technologies, non-AI technologies do not involve advanced algorithms or machine learning capabilities but still play a crucial role in facilitating access to information, enhancing communication, and supporting educational activities. These technologies provide versatile platforms for m-learning (mobile learning), e-learning, and blended learning environments, offering substantial benefits to educational processes.

\subsubsection{Laptops and Tablets in Education}

Laptops and tablets have become staple tools in educational institutions, widely recognised for their ability to enhance learning experiences by providing both students and educators with flexible access to educational resources. Research highlights that laptops are often perceived as more effective than mobile phones for educational purposes due to their larger screens and more robust computing capabilities, which are better suited for creating and managing digital content (Şad \& Göktaş, 2013). These devices support a wide range of educational software and applications, enabling interactive learning experiences that can accommodate diverse learning styles.

\subsubsection{M-learning}

Mobile learning (M-learning), which utilises mobile devices to deliver educational content, allows learning to occur anywhere and anytime, thereby increasing the accessibility of education. Mobile phones, in particular, are highly portable, making them excellent tools for learning on the go. Despite their smaller screens, they are beneficial for tasks that require quick access to information and for communication-related educational activities (Şad \& Göktaş, 2013). However, studies have shown that while mobile phones are highly valued for their portability, they are less favoured for more intensive educational tasks compared to laptops and tablets due to their smaller screen sizes and more limited input capabilities (Kukulska-Hulme, 2007).

\begin{figure*}[htp]
    \centering
    \includegraphics[width=10cm]{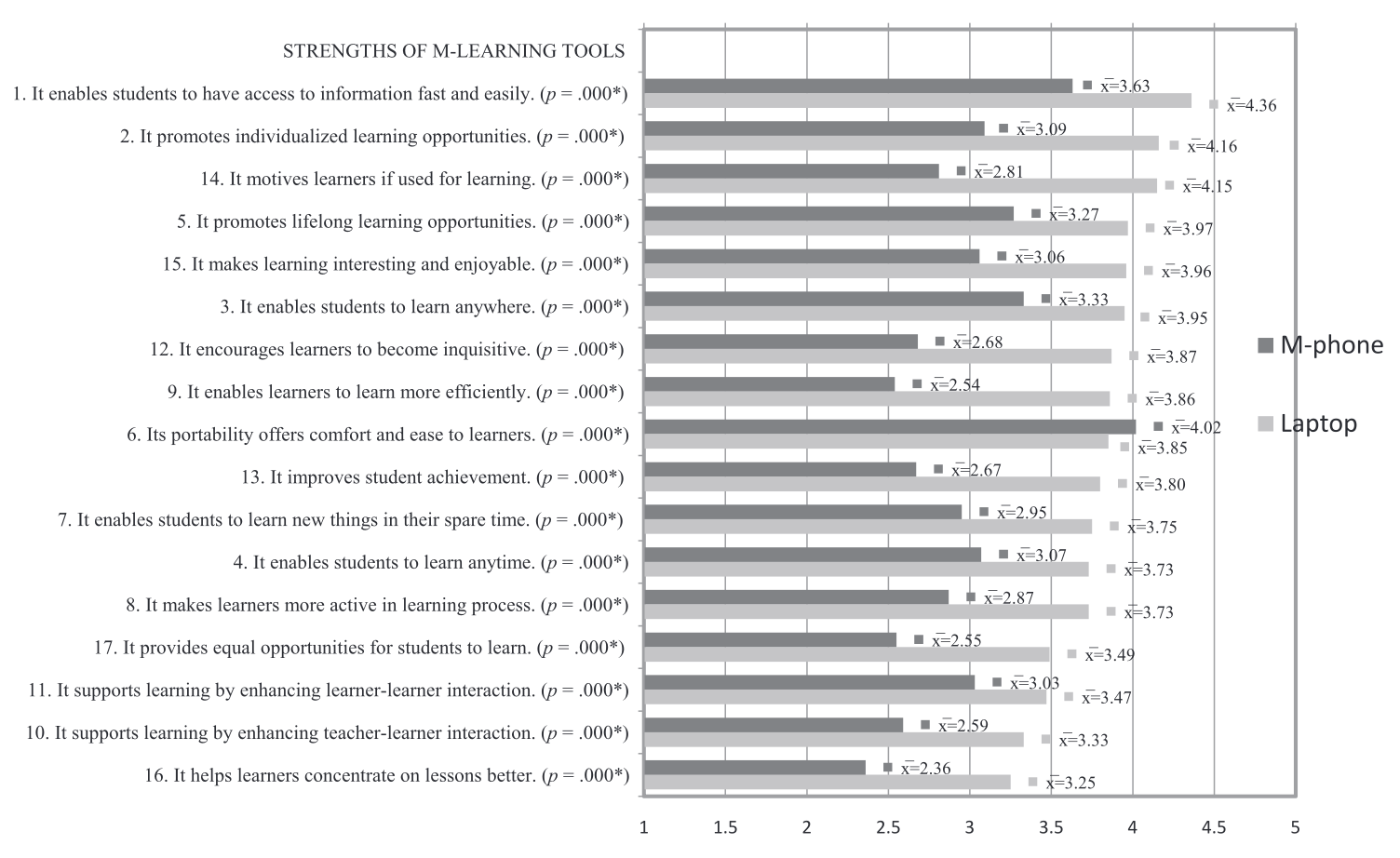}
    \caption{Strengths of M-Learning (Şad \& Göktaş, 2013)}
    \label{fig:sm}
\end{figure*}

\subsubsection{Research Outcomes on Non-AI Technology Usage}

Extensive research has explored the impact of non-AI technology in education. One significant finding is that while these technologies enhance educational accessibility and flexibility, they do not inherently guarantee improved educational outcomes. The effectiveness of these tools largely depends on how they are integrated into the curriculum and whether they are used to complement traditional educational methods effectively (Bingimlas, 2009). For example, the integration of laptops in the classroom has been shown to improve research and writing skills, but these benefits are contingent upon effective pedagogical strategies and adequate training for both teachers and students (Hew \& Brush, 2007).

\subsubsection{Challenges and Considerations}

Despite the advantages, the integration of non-AI technology in education is not without challenges. One of the primary concerns is the digital divide, which can exacerbate educational inequalities if not adequately addressed. Students without reliable access to digital devices or the internet may find themselves at a disadvantage, highlighting the need for policies that ensure equitable access to technology (Warschauer, 2006). Furthermore, there is the issue of distraction, as devices like mobile phones can lead to off-task behaviour if not managed properly within the educational setting (Rosen, Lim, Carrier, \& Cheever, 2011).

\begin{figure*}[htp]
    \centering
    \includegraphics[width=10cm]{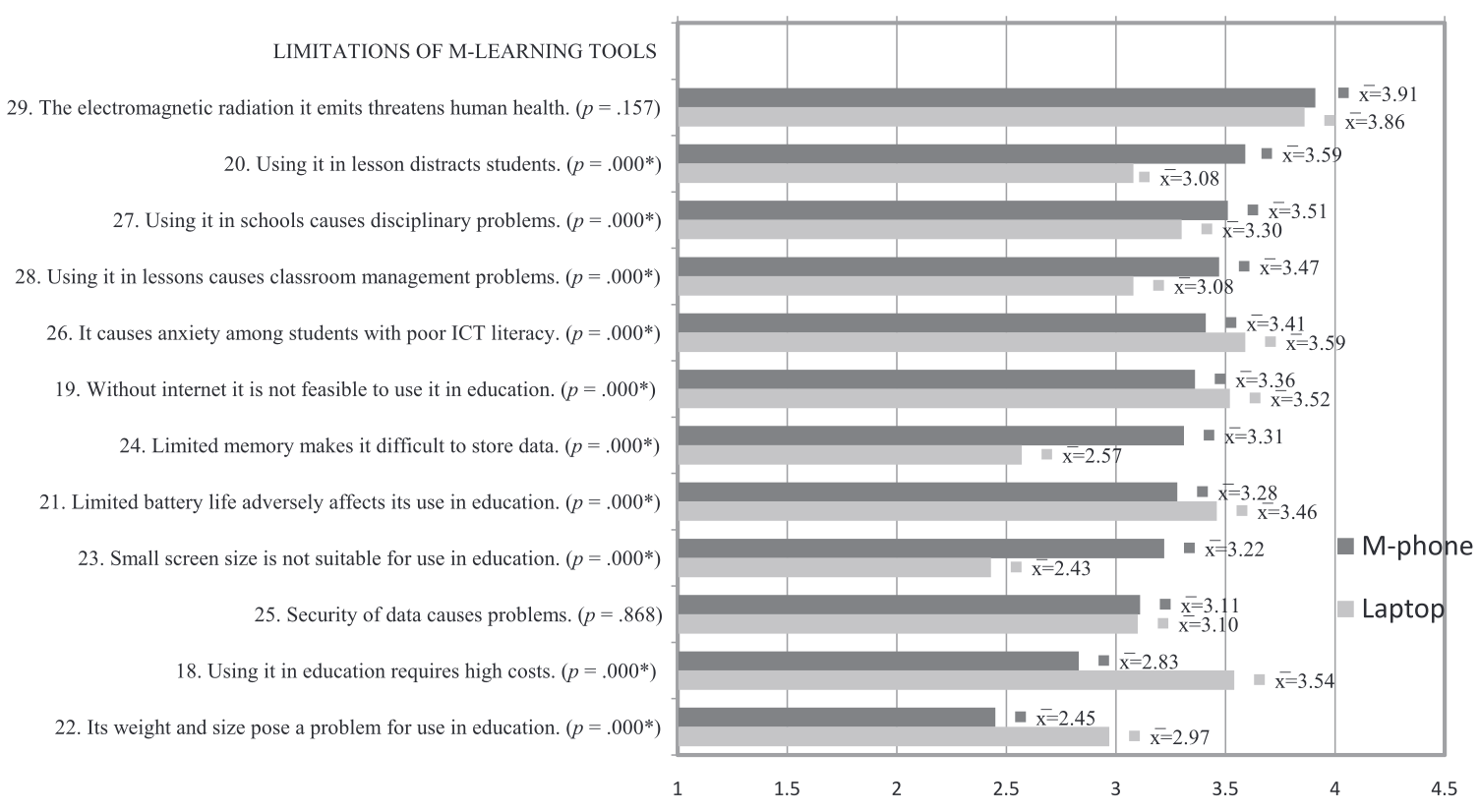}
    \caption{Limitations of M-Learning (Şad \& Göktaş, 2013)}
    \label{fig:lm}
\end{figure*}

In light of these challenges, the Claxton Future of Teaching framework provides a useful perspective on how to effectively incorporate technology in education while addressing issues like the digital divide and potential distractions. This framework emphasises the importance of teaching students not just content but also how to learn and adapt in a technology-rich environment. It advocates for the development of 'learning power', which includes resilience, curiosity, creativity, and critical thinking skills that are crucial for navigating and harnessing the benefits of digital tools (Claxton, 2002).

By focusing on these competencies, educators can create a more inclusive and effective learning environment that prepares students for a future where technology is ubiquitous. The framework suggests practical strategies for integrating technology into the classroom in ways that enhance learning without causing distraction. This includes structured guidance on the use of technology for research and collaboration, as well as clear policies on device usage to minimise off-task behaviour. Additionally, it highlights the importance of teacher training in digital literacy, ensuring that educators are equipped not only to use technology effectively but also to teach their students how to use it responsibly and effectively in their learning processes.

The integration of non-AI technology in education offers significant benefits by enhancing the accessibility, engagement, and flexibility of educational experiences. However, for these technologies to be truly effective, they must be integrated thoughtfully, with attention to pedagogical approaches and equitable access. Future research should continue to explore innovative ways to harness these technologies to support diverse educational needs and to overcome the challenges associated with their use in educational settings.

\subsection{AI-Driven Pedagogy}

\subsubsection{Current Implementations and Developments}

Artificial Intelligence (AI) is profoundly transforming the educational landscape by introducing personalised and adaptive learning systems that cater specifically to the needs of individual learners. This shift not only enhances student engagement but also optimises educational outcomes by dynamically adjusting teaching methods and materials (Zawacki-Richter, O., et al., 2019; Holmes, W., et al., 2021). The adoption of advanced AI technologies in educational contexts marks a significant leap towards creating more dynamic and responsive learning environments. By integrating machine learning algorithms and natural language processing, educational systems can develop intelligent tutoring systems and adaptive learning platforms that are more attuned to the individual needs of students. These AI-driven systems are particularly adept at processing vast amounts of student data, from which they can extract meaningful patterns and insights. This capability allows for the provision of customised feedback and the development of tailored instructional strategies, significantly enhancing the learning experience.

Machine learning algorithms, a core component of these AI systems, enable the automated analysis of student performance data. This analysis helps in identifying students' learning habits, strengths, and areas needing improvement, allowing for the adjustment of educational content to match their learning pace and style. This personalisation of learning materials and approaches aims to maximise educational outcomes by catering more directly to the learning preferences and requirements of each student (Zhou, M., \& Brown, D., 2020).

Moreover, natural language processing (NLP) technologies play a critical role in facilitating interactions between students and AI educational platforms. NLP allows these systems to understand and process human language, enabling them to communicate effectively with students. This interaction enhances the accessibility and usability of learning platforms, making them more engaging and easier to navigate for students. NLP technologies can also assist in the creation of new educational content, such as generating practice test questions or summarising key information from textbooks or databases (Johnson, L. et al., 2020).

\begin{figure}[htp]
    \centering
    \includegraphics[width=10cm]{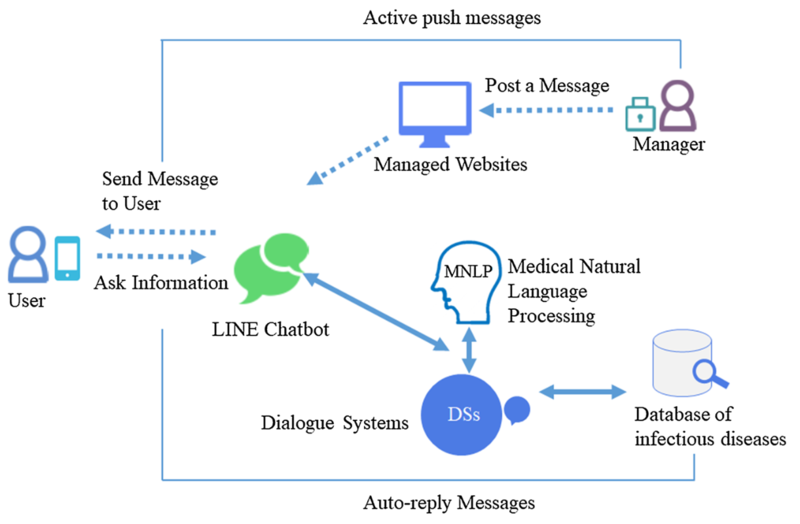}
    \caption{Sample NLP Dialogue System (Lee, Hwang, \& Chen (2022))}
    \label{fig:sm}
\end{figure}

Together, these technologies not only streamline various aspects of the educational process but also significantly enhance the efficacy and efficiency of learning by providing real-time, data-driven insights. Educators can utilise these insights to refine their teaching strategies and interventions, thus further enhancing the quality of education provided (Baker, R. S., \& Smith, L., 2019).

In essence, the integration of AI technologies into educational settings represents a transformative development that promises to make learning more personalised, efficient, and effective. As these technologies continue to evolve, their potential to revolutionise the educational landscape continues to expand, pointing towards a future where education is deeply personalised and widely accessible.

\subsubsection{AI in Educational Assessment and Management}

Furthermore, AI tools are increasingly being adopted within educational institutions to automate administrative tasks and enhance assessment processes. This automation empowers educators by allowing for the real-time analysis of student responses. AI systems can promptly provide feedback to students, and they enable educators to quickly identify and remedy learning difficulties, significantly improving educational outcomes (Zhou, M., \& Brown, D., 2020). The ability of AI to handle large datasets and provide instantaneous feedback not only streamlines assessment processes but also supports a more dynamic and responsive educational environment, which often leads to improved results. (Zhou, M., \& Brown, D., 2020).

\subsubsection{Enhancing Higher-Order Thinking Skills}

Beyond improving administrative efficiency, AI continues to evolve and expand within educational systems by offering promising enhancements to educational practices. Further enhancing the educational landscape, AI applications crucially support the development of critical thinking and problem-solving skills. Through AI-driven analytics, these systems can effectively analyse learning patterns and predict student performance. This information is invaluable for educators as it guides the development of pedagogical strategies that foster higher-order cognitive skills. Such strategies are essential for academic success and prepare students to tackle complex problems in their future careers, thus bridging the gap between educational settings and professional demands (Johnson, L. et al., 2020; Stevens, A., \& Marden, P., 2021).

\subsubsection{Challenges and Ethical Considerations}

However, the integration of AI into educational practices is not devoid of challenges. Ethical concerns pose significant obstacles, and to navigate these issues, it is imperative that educational policymakers and practitioners collaborate to establish protections, ethical guidelines, and transparency guidelines promoting safety and equity in educational opportunities. Such efforts are crucial to ensuring that AI is integrated into educational contexts ethically and responsibly, fostering an environment where all students can benefit from these advanced technologies (Bostrom, N., \& Yudkowsky, E., 2014; Weller, M., 2021).

One of the foremost ethical issues with the use of AI in education is the protection of student data. AI systems often require access to large datasets, including sensitive personal information, to function optimally. This raises significant concerns about data privacy and security, especially regarding how data is collected, stored, used, and shared. To address these concerns, it is crucial for educational institutions to implement stringent data protection measures. This includes encrypting student data, ensuring secure data storage and transmission, and establishing clear policies regarding data access and use. Additionally, there should be transparency about data practices with students and parents, ensuring they are informed and have control over their personal information.

Another critical ethical issue is the potential for AI systems to exhibit and even amplify existing biases. Algorithmic bias can occur when AI systems make decisions based on skewed data or biased algorithms, leading to unfair outcomes for certain groups of students. To mitigate this risk, AI systems used in education must be designed and regularly audited for fairness and accuracy. This involves diverse and inclusive data training sets and algorithms that are regularly reviewed and updated to eliminate biases. Furthermore, there should be mechanisms to challenge and correct AI decisions, ensuring that human oversight is a continuous aspect of the AI implementation process.

The deployment of AI in education also risks widening the digital divide, particularly affecting students from underprivileged backgrounds who may have limited access to digital tools and technologies. To ensure equity in educational opportunities, it is essential to make AI tools accessible to all students. This might involve providing subsidised or free access to necessary hardware and software, investing in broadband infrastructure to ensure reliable internet access, and offering training programs for students and educators to effectively utilise AI technologies.

To navigate these ethical challenges effectively, a collaborative approach involving educational policymakers, practitioners, and technology experts is vital. This collaborative framework should focus on establishing robust ethical guidelines and standards for AI use in education, emphasising privacy, fairness, and accessibility. Continuous monitoring and evaluation should be part of this framework, with the flexibility to adapt policies and practices as AI technologies and their applications in education evolve.

\subsubsection{Looking Ahead: The Future of AI in Education}

The trajectory of AI in education is gearing towards more sophisticated, intuitive, and collaborative tools, promising to significantly enhance the interaction between students and the educational content they engage with. These advancements are expected to not only make learning more engaging but also more effective, tailoring educational experiences to individual needs and preferences more seamlessly than ever before. As AI tools become more integrated into educational systems, they are poised to transform traditional learning environments into dynamic, interactive spaces that encourage deeper understanding and greater student involvement (Reid, P., \& Lai, K., 2021).

However, the rapid development of such technologies also necessitates the creation of robust frameworks to govern their use. This is critical to ensure that the deployment of AI technologies in education is done ethically and equitably. Effective governance frameworks will need to address key issues such as privacy, bias, accessibility, and the impact of AI on the educational workforce. Ensuring that AI innovations benefit all participants equitably will require continuous assessment and adaptation of these frameworks as the technology evolves (Weller, M., 2021).

Moreover, as AI technologies become more prevalent in educational settings, there will be an increasing need for educators to be trained not just in the use of these technologies, but also in understanding their implications. This includes training on the potential biases inherent in AI systems, the ethical use of student data, and strategies to integrate AI tools in a way that complements traditional teaching methods rather than replacing them.

The future of AI in education is not just about technological advancement but also about ensuring that these advancements lead to fairer, more inclusive, and more effective educational experiences. As such, the ongoing dialogue between technologists, educators, policymakers, and other stakeholders will be crucial in shaping an educational landscape that harnesses the potential of AI while mitigating its risks.

\section{Introduction to TriDENT: Tripartite Developmental Education with Neural Technology}

The evolving landscape of educational technology presents both challenges and opportunities to enhance the learning experience. Traditional approaches to education have emphasised foundational knowledge but often lack personalised learning pathways that cater to individual student needs. Meanwhile, technological advancements, including non-AI and AI-driven systems, offer new ways to enrich and tailor education. TriDENT, or Tripartite Developmental Education with Neural Technology, is an educational system proposed in this research consciously designed to merge the best of these worlds by integrating the principles of classical education – Logic (Understanding), Grammar (Knowledge), and Rhetoric (Wisdom) – with modern technological innovations. This system aims to leverage the strengths of traditional, non-AI, and AI-driven educational methodologies to create a holistic, adaptive, and deeply engaging learning environment.

\subsection{Educational Framework of TriDENT}
TriDENT is structured around the classical trivium, which consists of three stages: Grammar, Logic, and Rhetoric. Each stage is aligned with specific technological tools to maximise learning efficiency and effectiveness:

\begin{enumerate}
    \item Grammar (Knowledge): This initial phase focuses on the acquisition of foundational knowledge. In TriDENT, traditional classroom settings are enhanced with non-AI technological supports such as digital textbooks, interactive media, and online databases. These tools provide students with access to a wealth of information and facilitate the efficient transmission of factual content, as suggested by Lee, Hwang, \& Chen (2022) who emphasise the importance of integrating digital tools to improve access to information and collaborative learning.

    \item Logic (Understanding): The second phase involves developing critical thinking and analytical skills. Here, AI-driven educational technologies like adaptive learning systems and intelligent tutoring come into play. These AI systems analyse individual learning patterns and tailor instructional strategies to each student’s needs, enhancing the understanding of complex concepts (Johnson, L. et al., 2020). By applying machine learning algorithms, TriDENT can provide personalised learning experiences that adapt in real-time to the evolving capabilities of each student, as outlined by Zhou, M., \& Brown, D. (2020).

    \item Rhetoric (Wisdom): In the final phase, students apply their knowledge and understanding to express ideas creatively and persuasively. AI tools such as natural language processing can aid in the development of writing and presentation skills, providing feedback that helps refine argumentation and rhetorical skills. This stage integrates AI to support creative expression, allowing for the simulation of real-world scenarios where students can practice and hone their rhetorical abilities in a controlled, feedback-rich environment.

\end{enumerate}

\subsection{Implementation and Technological Integration}

Implementing the TriDENT system requires a comprehensive and stratified approach to technology integration, ensuring that each educational phase is enhanced by the most appropriate technological tools. This integration aims to create a seamless educational experience that supports both teachers and students through various stages of learning. Here’s how TriDENT can be implemented effectively:

A robust IT infrastructure is foundational to the successful deployment of the TriDENT system. This infrastructure must include high-speed internet access to facilitate seamless communication and resource sharing. Cloud computing resources are essential for providing scalable and flexible access to computing power and data storage, which are crucial for handling the extensive data processing required by AI systems. Moreover, secure data storage solutions must be implemented to protect sensitive educational data. This infrastructure will support the diverse technologies employed in TriDENT, ensuring that they are reliable and continuously available for educational purposes. To fully harness the capabilities of TriDENT, educators must be well-versed not only in traditional teaching methods but also in the use of advanced AI tools. Professional development programs should be established to train educators on the effective use of AI in the classroom. These programs should cover various aspects, including the ethical use of AI, strategies for data security, and the pedagogical integration of technology into teaching. By focusing on these areas, educators will be better equipped to utilise AI tools in a way that enhances learning while maintaining a focus on sound educational principles.

Each stage of the TriDENT educational process should be carefully aligned with specific technological tools that enhance learning:

\begin{itemize}

    \item Grammar Stage (Knowledge Acquisition): In this initial phase, traditional classroom settings should be enhanced with non-AI technological supports such as digital textbooks and interactive media. These tools augment foundational knowledge acquisition by providing students with a broader range of learning materials and interactive content, making the learning process more engaging and comprehensive.

    \item Logic Stage (Critical Thinking Development): AI-driven educational technologies become particularly valuable in this phase. Adaptive learning systems and intelligent tutoring can be integrated to personalise instruction and provide feedback based on individual student performance. This personalisation helps develop critical thinking skills by allowing students to progress at their own pace and focus on areas that need improvement.

    \item Rhetoric Stage (Effective Communication): At this stage, technologies such as natural language processing can support the development of students' communication skills. AI tools can assist in refining students' written and oral expression, providing feedback that enhances their ability to articulate thoughts clearly and persuasively.
    
\end{itemize}

\subsection{Challenges and Future Directions}
While TriDENT promises to revolutionise educational practices, several challenges need addressing, including the digital divide, which could limit access to AI-enhanced learning tools. Proactive measures, such as providing subsidised technology access and enhancing digital literacy, are essential to ensure all students benefit from TriDENT’s offerings.

The deployment of AI technologies within the TriDENT framework must adhere to strict ethical standards to ensure the protection of student privacy and the promotion of fairness. It is essential to incorporate algorithms designed to minimise bias and promote equity in educational outcomes. This involves not only selecting and designing AI systems with these considerations in mind but also continuously monitoring and adjusting them to prevent discriminatory practices. Ethical guidelines should be developed and adhered to, ensuring that all students benefit equally from AI-enhanced education.

TriDENT represents a forward-thinking approach to education that integrates the time-tested educational framework of the trivium with cutting-edge technology. By doing so, it seeks to prepare students not only academically but also for the challenges of a digital world, ensuring they are knowledgeable, thoughtful, and wise. As technology continues to advance, TriDENT will evolve, continually integrating new tools and techniques to remain at the forefront of educational innovation.

\section{Analysis of TriDENT}

TriDENT, an innovative educational framework, melds the classical education theory of the trivium – Grammar, Logic, and Rhetoric – with cutting-edge non-AI and AI-driven technologies. This integration is designed to elevate each stage of learning, enhancing knowledge acquisition, critical thinking, and effective communication. The trivium’s structured approach to education, which progresses students through foundational learning to sophisticated expression, aligns perfectly with the capabilities of modern educational technologies to provide a comprehensive and enriched learning experience.

In practice, TriDENT leverages non-AI technology in the Grammar stage to augment foundational knowledge. Digital textbooks and interactive media provide a dynamic platform for content delivery, aligning with the trivium’s focus on rigorous knowledge acquisition. As students advance to the Logic stage, AI-driven analytics offer personalised learning experiences that foster deeper understanding and critical analysis. These systems adapt to individual learning patterns, enhancing the educational process by tailoring content and feedback to the needs of each student. In the final Rhetoric stage, AI tools like natural language processing empower students to articulate their ideas creatively and persuasively, fulfilling the trivium’s emphasis on eloquent and effective communication.

\subsection{Challenges and Limitations}

Despite its potential, TriDENT faces several challenges. The reliance on advanced AI technologies risks exacerbating the digital divide, potentially sidelining underprivileged students who lack access to necessary technological resources. Furthermore, the effectiveness of TriDENT is highly dependent on the integrity and fairness of the AI algorithms employed. Poorly managed or biased algorithms could lead to discriminatory practices within the educational system. There is also a concern that an overemphasis on technology could undermine the role of educators, diminishing the invaluable human element in teaching and potentially neglecting the development of interpersonal skills and emotional intelligence in students.

To fully realise TriDENT’s potential, extensive research is required to address its current limitations and to refine its application. Initial research should focus on assessing the accessibility of AI technologies, ensuring that these tools can be equitably distributed and are usable in diverse educational settings. Studies should also evaluate the efficacy of AI-driven personalisation techniques, measuring their impact on student learning outcomes across different demographics and learning styles. Empirical research is needed to explore the integration of AI within the trivium framework. This includes longitudinal studies to assess the long-term effects of AI-enhanced education on critical thinking and communication skills. Additionally, research into the ethical deployment of AI in education should be prioritised to develop guidelines and best practices for maintaining student privacy, ensuring data security, and preventing algorithmic bias.

\subsection{Critics and Rebuttals}

Critics of integrating AI in educational settings raise significant concerns that need thorough consideration to ensure the effective and ethical use of technology. One major concern is data privacy; integrating AI involves processing large volumes of personal data from students, which poses risks such as unauthorised data access and breaches. Another worry is that AI could diminish the crucial human interaction in learning environments, as automated systems might replace the valuable face-to-face engagements that are integral to student development. Additionally, there is a legitimate fear that AI systems could inadvertently perpetuate existing societal biases if they are trained on historically biased data or implemented without consideration of their broader impacts.

To address these concerns, the TriDENT framework proposes the development and enforcement of robust ethical frameworks alongside continuous oversight mechanisms. Such frameworks must prioritise transparency and accountability in AI applications within educational systems. This involves not only the clear documentation and communication of how AI systems operate and are applied in educational contexts but also includes regular audits of these systems. These audits should assess AI algorithms for fairness and accuracy, ensuring they do not perpetuate biases or inaccuracies that could harm student outcomes. Moreover, to maintain essential human interaction within learning environments, TriDENT should advocate for a blended learning model. This model would strategically integrate AI tools to enhance, rather than replace, traditional teaching methods. AI should be seen as a supplement that enhances educational experiences and outcomes through personalised learning and efficient administrative processes, not as a replacement for the human elements of teaching that are critical for developing students' social skills and emotional intelligence. Implementing AI in this thoughtful and supportive manner can help mitigate critics' concerns and pave the way for more accepted and effective use of technology in education. This approach ensures that AI serves as an empowering tool for both teachers and students, enhancing educational practices while preserving the core values of traditional education.
\subsection{Realisation of Ideals and Recommendations}

The ideals of TriDENT – integrating classical education with cutting-edge technology – are ambitious but achievable with the right measures. Ensuring equitable access to technology, providing comprehensive training for educators, and establishing strong ethical guidelines are crucial steps towards realising these ideals. Research indicates that continuous development and assessment of AI applications in education will be necessary to adapt to the evolving educational needs and technological advancements (Reid, P., \& Lai, K., 2021).

\subsection{Future Directions and Theoretical Considerations}

Moving forward, TriDENT should focus on developing more inclusive technologies that can bridge the digital divide. There is a need for further theoretical and empirical research into how AI can be tailored to enhance not just academic performance but also social and emotional learning. Researchers should also explore the long-term impacts of AI in education, particularly concerning student autonomy and motivation. While TriDENT holds an overall great promise in revolutionising educational practices through the integration of the trivium and modern technology, it requires careful implementation and continuous evaluation to ensure that it truly enhances educational outcomes in an equitable and ethical manner.

Looking ahead, TriDENT should contribute to the innovation of more inclusive technologies that bridge the digital divide and enhance both academic and emotional intelligence. This requires the development of AI tools that are not only academically robust but also sensitive to the social and emotional aspects of learning. Recommendations for policymakers and educational leaders include investing in infrastructure that supports widespread technology access, creating professional development programs for educators on the use of AI, and establishing a regulatory framework that governs the ethical use of AI in education. The success of this framework hinges on thoughtful implementation, rigorous research, and a steadfast commitment to equity and ethics. Continuous evaluation and adaptation will be key to ensuring that TriDENT not only meets the current educational demands but also adapts to future challenges and opportunities.


\begin{thebibliography}{9}
\bibitem{Laurillard2002}
Laurillard, D. (2002). \textit{Rethinking University Teaching: A Conversational Framework for the Effective Use of Learning Technologies}.

\bibitem{Alexander2004}
Alexander, R. (2004). \textit{Towards Dialogic Teaching: Rethinking Classroom Talk}.

\bibitem{Bates2015}
Bates, A. W. (2015). \textit{Teaching in a Digital Age: Guidelines for Designing Teaching and Learning}.

\bibitem{Means2010}
Means, B. (2010). Technology and education change: Focus on student learning.

\bibitem{Hirsch1987}
Hirsch, E. D. (1987). \textit{Cultural Literacy: What Every American Needs to Know}. Houghton Mifflin.


\bibitem{Mayer2009}
Mayer, R. E. (2009). \textit{Multimedia Learning}.

\bibitem{Voogt2012}
Voogt, J., \& Roblin, N. P. (2012). A comparative analysis of international frameworks for 21st-century competences: Implications for national curriculum policies. \textit{Journal of Curriculum Studies}.

\bibitem{Selwyn2016}
Selwyn, N. (2016). \textit{Education and Technology: Key Issues and Debates}.

\bibitem{Zhao2012}
Zhao, Y. (2012). \textit{World Class Learners: Educating Creative and Entrepreneurial Students}.

\bibitem{Tapscott1998}
Tapscott, D. (1998). \textit{Growing up digital: The rise of the Net Generation}.

\bibitem{Prensky2001}
Prensky, M. (2001). Digital natives, digital immigrants. \textit{On the Horizon}, MCB University Press.

\bibitem{Oblinger2005}
Oblinger, D. G., \& Oblinger, J. L. (Eds.). (2005). \textit{Educating the net generation}. Educause.

\bibitem{Williamson2021}
Williamson, B. (2021). The future of teaching and the myths of educational reform.

\bibitem{Chen2020}
Chen, Lijia, Chen, Pingping, \& Lin, Zhijian. (2020). Artificial intelligence in education: A review. \textit{IEEE Access}, 8, pp. 75264-75278.

\bibitem{Aggarwal2023}
Aggarwal, Deepshikha. (2023). Integration of innovative technological developments and AI with education for an adaptive learning pedagogy. \textit{China Petroleum Processing and Petrochemical Technology}, 23(2).

\bibitem{Lee2022}
Lee, Yen-Fen, Hwang, Gwo-Jen, \& Chen, Pei-Ying. (2022). Impacts of an AI-based chatbot on college students’ after-class review, academic performance, self-efficacy, learning attitude, and motivation. \textit{Educational Technology Research and Development}, 70(5), pp. 1843-1865.

\bibitem{Fu2020}
Fu, Shixuan, Gu, Huimin, \& Yang, Bo. (2020). The affordances of AI-enabled automatic scoring applications on learners’ continuous learning intention: An empirical study in China. \textit{British Journal of Educational Technology}, 51(5), pp. 1674-1692.

\bibitem{ZawackiRichter2019}
Zawacki-Richter, Olaf, Marín, Victoria I., Bond, Melissa, \& Gouverneur, Franziska. (2019). Systematic review of research on artificial intelligence applications in higher education – where are the educators? \textit{International Journal of Educational Technology in Higher Education}, 16, article no. 39.

\bibitem{Baker2019}
Baker, R. S., \& Smith, L. (2019). Insights into engagement and learning through educational data mining and learning analytics.

\bibitem{Bostrom2014}
Bostrom, N., \& Yudkowsky, E. (2014). The ethics of artificial intelligence.

\bibitem{Woolf2010}
Woolf, B. P. (2010). Building intelligent interactive tutors: Student-centered strategies for revolutionizing e-learning.

\bibitem{Zhou2020}
Zhou, M., \& Brown, D. (2020). Machine learning in educational settings.

\bibitem{Reid2021}
Reid, P., \& Lai, K. (2021). Future directions for teaching and learning with AI.

\bibitem{Stevens2021}
Stevens, A., \& Marden, P. (2021). The impact of AI on higher education.

\bibitem{Weller2021}
Weller, M. (2021). 21st Century Education: Opportunities and Challenges.

\bibitem{ZawackiRichter2019bis}
Zawacki-Richter, O., et al. (2019). Systematic review of research on artificial intelligence applications in higher education – where are the educators?
\end{thebibliography}
\end{document}